\documentclass[twocolumn,prl,preprintnumbers,amsmath,amssymb,superscriptaddress]{revtex4}

\usepackage{graphicx}% Include figure files
\usepackage{dcolumn}% Align table columns on decimal point
\usepackage{bm}% bold math
\usepackage{array}
\usepackage{mathrsfs}
\usepackage{physics}
\usepackage{times}
\usepackage{float}
\usepackage{booktabs}
\usepackage{autobreak}
\usepackage{afterpage}
\graphicspath{{figures/}}
\usepackage[colorlinks=true,linktocpage=true,linkcolor=blue,citecolor=blue]{hyperref}

\graphicspath{{figures/}}
\newcolumntype{I}{!{\vrule width 1pt}}
\newlength\savedwidth

\newlength\savewidth

\begin{document}

\title{ Configuration entropy of $\Upsilon(1S)$ state in strong coupling plasma}

\author{Yan-Qing Zhao}
\email{zhaoyanqing@mails.ccnu.edu.cn}
\affiliation{Institute of Particle Physics and Key Laboratory of Quark and Lepton Physics (MOS),
Central China Normal University, Wuhan 430079, China}

\author{Defu Hou}
\thanks{Corresponding author}
\email{houdf@mail.ccnu.edu.cn}
\affiliation{Institute of Particle Physics and Key Laboratory of Quark and Lepton Physics (MOS),
Central China Normal University, Wuhan 430079, China}

\date{\today}

%%%%%%%%%%%%%%%%
\begin{abstract}
To better understand the effects of strong coupling and QCD at high temperature in QGP, by using holographic model, we investigate the dissociation effect of bottomonium under the higher-order curvature corrections to the supergravity action corresponding to the corrections of large N expansion of boundary CFT in the side of field theory. The results show that effective potential is not a good physical quantity to estimate the dissociation strength of bottomonium in the case of finite wave number and considering the higher-order curvature corrections. Therefore, we calculate the quasinormal spectra(QNMs) and the differential configuration entropy(DCE). It is found that the dissociation effect is stronger for the stronger coupling.
\end{abstract}
%\pacs{}
%\showpacs{}
\maketitle

%%%%%%%%%%Section1%%%%%%%%%%
\section{\label{sec:01_intro} Introduction}

To investigate the formation and evolution of quark gluon plasma(QGP), the study of heavy quarkonium($J/\psi$ and $\Upsilon(1S)$ mainly) production rate plays an irreplaceable role in the process. Because the QGP matter can not be observed directly. It is well-known that heavy vector mesons may still survive when light mesons have dissolved, that is to say, the physical system has undergone the deconfinement phase transition.  So far, there are many physical quantities and methods that can be used to study the properties of heavy flavors. For example, the real potential~\cite{Finazzo:2014rca,Kim:2008ax,Hou:2007uk,BoschiFilho:2006pe,Feng:2019boe}, imaginary potential~\cite{Zhao:2019tjq,Feng:2019boe,Zhang:2019qhm,Fadafan:2015kma,Finazzo:2014rca}, thermal width~\cite{Feng:2019boe,Braga:2016oem,Fadafan:2015kma,Finazzo:2014rca,Fadafan:2013coa,Zhao:2019tjq}, separation length~\cite{Feng:2019boe,Liu:2006nn,Liu:2006he}, thermal spectral functions~\cite{Braga:2017apr,Zhao:2021ogc} of heavy quarkonium and so on.

In recent years a great deal of interest has been shown in using the configuration entropy(CE) to study the stability of $Q\bar{Q}$ pair, which is put forward by Gleiser and Stamatopoulos in refs.~\cite{Gleiser:2011di,Gleiser:2012tu,Gleiser:2013mga}. Configuration entropy is closely related to quasinormal modes. In order to get the configuration entropy, one must look for complex frequencies of the heavy quarkonium by solving the equation of motion of quasi-states with related boundary condition. The real part of the complex frequency are associated with the thermal mass and the imaginary part with the thermal width of heavy quarkonium. Many research results show that the more stable the system, the smaller the CE value~\cite{Gleiser:2015rwa,Bernardini:2016hvx,Bernardini:2016qit,Braga:2017fsb,Braga:2018fyc,Bernardini:2018uuy,Ferreira:2019inu,Alves:2014ksa,Correa:2015vka,Gleiser:2011di,Correa:2015qma,Braga:2016wzx,Braga:2017fsb,Karapetyan:2017edu,Karapetyan:2018oye,Karapetyan:2018yhm,Bazeia:2018uyg,Lee:2017ero,Colangelo:2018mrt,Lee:2018zmp,Bernardini:2019stn,Karapetyan:2019ran,Karapetyan:2019fst,Gleiser:2020zaj,Cruz:2019kwh,Cruz:2018qby,Bhattacharjee:2020gzl}. In ref.~\cite{Braga:2020hhs}, configuration entropy of charmonium is calculated by considering the influence of magnetic fields. In ref.~\cite{Braga:2020myi}, they develop the configuration entropy for quarkonium in a finite density plasma. The definition of configuration entropy originated from Shannon information entropy~\cite{Shannon:1948zz} that gives the information contained in the variable $a$, which may have the discrete value $a_n$ and the probabilities $p_n$,
\begin{equation}\label{eq1}
  S[f]=-\sum_{n}p_n\log p_n,
\end{equation}
where $p_n$ satisfies the normalization condition: $\sum_{n}p_n=1$. For successive quantity, Shannon information entropy reads
\begin{equation}\label{eq2}
 S[f]=-\int d^dx \epsilon(\vec{x})\log\epsilon(\vec{x}),
\end{equation}
where modal fraction,
\begin{equation}\label{eq3}
  \epsilon(\vec{x})=\frac{\vert{\rho(\vec{x})}\vert^2}{\int d^dx\vert{\rho(\vec{x})}\vert^2},
\end{equation}
satisfies normalization condition $\int d^dr\epsilon(\vec{x})=1$. The normalizable function in coordinate space $\rho(\vec{x})$ needs to be transformed into momentum space by Fourier transformation for the sake of calculating the configuration entropy,
\begin{equation}\label{eq4}
  \rho(\vec{k})=\frac{1}{(2\pi)^{d/2}}\int d^dx \rho(\vec{x})e^{-i\vec{k}\cdot\vec{x}}.
\end{equation}
Accordingly, the modal fraction is given by
\begin{equation}\label{eq5}
  \overline{\epsilon}(\vec{k})=\frac{\vert{\rho(\vec{k})}\vert^2}{\int d^dk\vert{\rho(\vec{k})}\vert^2}.
\end{equation}
Furthermore, the configuration entropy is introduced as follows
\begin{equation}\label{eq6}
  S_{CE}=-\int d^dk \overline{\epsilon}(\vec{k}) \log\overline{\epsilon}(\vec{k}).
\end{equation}
However, the value of CE may be negative for continuous variables. For convenience, another different modal fraction was introduced  in ~\cite{Gleiser:2018kbq}
\begin{equation}\label{eq7}
  \epsilon(\vec{k})=\frac{\vert{\rho(\vec{k})}\vert^2}{\vert{\rho(\vec{k})}\vert^2_{max}}.
\end{equation}
The resulting configuration entropy is called differential configuration entropy(DCE),
\begin{equation}\label{eq8}
 S_{DCE}=-\int d^dk \epsilon(\vec{k})\log\epsilon(\vec{k}).
\end{equation}

The study of quark-gluon plasma (QGP) is of great importance in understanding the fundamental properties of matter under extreme conditions. Holographic methods have provided a powerful tool for investigating QGP from a gravitational perspective. However, the effects of strong coupling and QCD at high temperature in QGP are still not well understood. To address this issue, we consider the high-order curvature corrections in the gravity theory when studying QGP using holographic methods. By doing so, we aim to gain a deeper insight into the behavior of QGP and uncover new and interesting phenomena, such as strong fluid properties and perturbation theories, which may have important implications for our understanding of QGP and the nature of matter at extreme conditions.

The main objective of the present study is to explore the quasinormal frequencies and the differential configuration entropy of heavy quarkonia in a hot dense matter by considering a class of bulk theories comprising higher-order curvature corrections to the supergravity action~\cite{Cai:2001dz}, which results in the conjectured viscosity bound~\cite{Kovtun:2004de} $\frac{\eta}{s}\geq\frac{1}{4\pi}$ to be violated. In the effective low-energy action of string theories, according to AdS/CFT dual, the higher-order curvature term can be regarded as the corrections of large N expansion of boundary CFT in the side of corresponding field theory. Using the holographic model for $\Upsilon(1S)$ in a hot dense plasma, the QNMs and the DCE of bottomonium are calculated. We will discuss the dissociation behavior of $Q\bar{Q}$ pair and possible implications on higher-order curvature corrections related effects.

  %%%%%%%%%%Section2%%%%%%%%%%

\section{\label{sec:02} The holographic model}
\begin{figure}
  \centering
  \includegraphics[width=7cm]{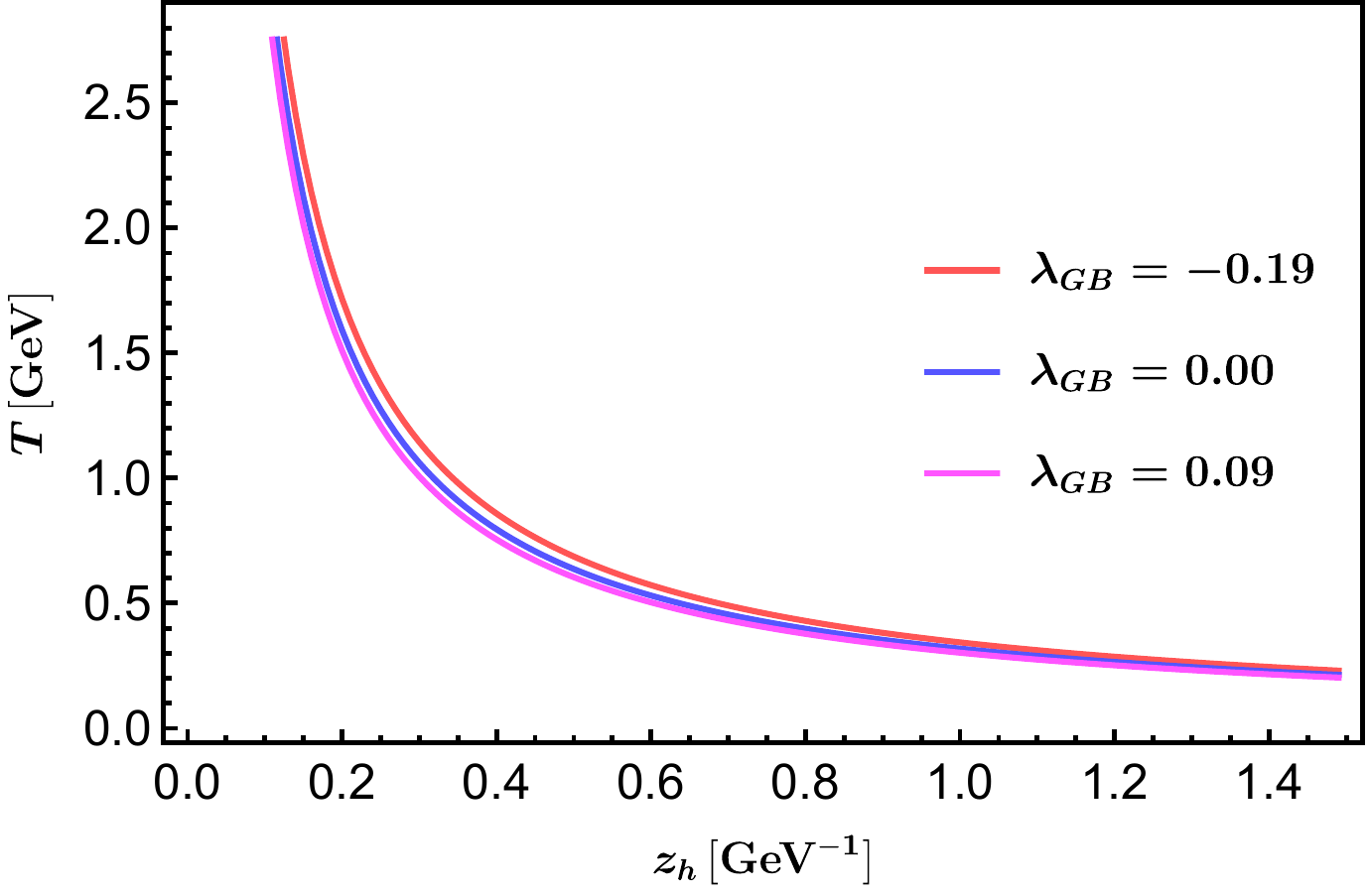}
  \caption{ The location of horizon $z_h$ dependence of black hole temperature $T$ with different constants $\lambda_{GB}$. The red curve represents $\lambda_{GB}=-0.19$; the blue curve is the result for $\lambda_{GB}=0$ and the magenta curve denotes $\lambda_{GB}=0.09$. }\label{fig1}
\end{figure}

For going deeply into higher-order curvature corrections, we take into account, a most general theory of gravity with quadratic powers of curvature in five-dimensional $(5D)$ spaces,Gauss-Bonnet(GB) gravity. The action is~\cite{Zwiebach:1985uq,Buchel:2009sk}
\begin{align}\label{eq9}
  S= \frac{1}{16\pi G_5}\int d^5x\sqrt{-G}[(\mathcal{R}+\frac{12}{R^2})+\frac{\lambda_{GB}R^2}{2}\mathcal{L}_{GB}],\\
  \mathcal{L}_{GB}=\mathcal{R}^2-4\mathcal{R}_{\mu\nu}\mathcal{R}^{\mu\nu}+\mathcal{R}_{\mu\nu\rho\sigma}\mathcal{R}^{\mu\nu\rho\sigma}.
\end{align}
Here $G_5$ is the five dimensional Newton constant, $R$ represents the radius of the asymptotic AdS spaces, $\mathcal{R}_{\mu\nu}$ labels the Ricci tensor and $\mu$ runs over 1 to 5, $\mathcal{R}_{\mu\nu\rho\sigma}$ denotes the Riemann tensor, $\mathcal{R}$ refers to the Ricci scalar, and $\lambda_{GB}$ is a coupling constant. The first polynomial gives the Einstein-Hilbert action and $\mathcal{L}_{GB}$ is the $\mathcal{R}^2$ corrections. The action exists a pinpoint black-brane solution, the metric can be written as~\cite{Cai:2001dz}
\begin{equation}\label{eq10}
   ds^2=\frac{R^2}{z^2}(-af_{GB}(z)dt^2+dx^2+\frac{1}{f_{GB}(z)}dz^2)
\end{equation}
with
\begin{equation}\label{eq11}
   a=\frac{1}{2}(1+\sqrt{1-4\lambda_{GB}}),
\end{equation}
\begin{equation}\label{eq12}
  f_{GB}(z)=\frac{1}{2\lambda_{GB}}(1-\sqrt{1-4\lambda_{GB}(1-\frac{z^4}{z_h^4})}).
\end{equation}
The black brane horizon $z_h$ is at $f_{GB}(z_h)=0$. Ref.~\cite{Cai:2001dz} points out that in five-dimensional spaces $\lambda_{GB}\le 9/100$, which can validly avoid causality violation at the boundary~\cite{Brigante:2008gz,Brigante:2007nu}. One can find that there is a vacuum AdS solution only when $\lambda_{GB}\le 0.25$ and exists a conformal field theory at the boundary. The Hawking temperature of the plasma can be obtained by the surface gravity, as follows
\begin{equation}\label{eq13}
  T=\abs{\frac{\kappa}{2\pi}}=\frac{\sqrt{a}}{\pi z_h},
\end{equation}
where $\kappa=\sqrt{-g_{tt}/g_{zz}}(\text{log}(\sqrt{-g_{tt}}))'|_{z=z_h}$ is the surface gravity. In Fig.\ref{fig1}, we show the behavior of black hole temperature $T$ as a function of the location of horizon $z_h$. As can be seen, increasing the location of horizon reduces the temperature, and the appearance of higher-order curvature corrections also diminishes the black hole temperature.

The constant $\lambda_{GB}$ and the ratio of the shear viscosity $\eta$ and the entropy density $s$ has the following relation~\cite{Brigante:2008gz,Brigante:2007nu,Kats:2007mq}
\begin{equation}\label{eq14}
  \frac{\eta}{s}=\frac{1}{4\pi}(1-4\lambda_{GB}).
\end{equation}
The viscosity bound, $\eta/s>1/4\pi$, is violated for $\lambda_{GB}>0$. The restraint $\lambda_{GB}\le 9/100$ indicates that $\eta/s\le 4/(25\pi)$. As can be seen from eq. (\ref{eq14}), for the same physical quantity, the behavior of the ratio $\eta/s$ and the constant $\lambda_{GB}$ is opposite. For convenience, in the following sections, we all use $\lambda_{GB}$ instead of $\eta/s$ to participate in the calculation.

Now, with the help of the phenomenological holographic model of~\cite{Braga:2019yeh}, we research the nature of heavy vector mesons in a hot dense plasma. We use a 5D dual vector field $A_m=(A_\mu, A_z)(\mu=0,1,2,3)$ to represent the heavy quarkonium. The action takes the following form
\begin{equation}\label{eq15}
  S_\text{matter}=\int d^4xdz\sqrt{-g}e^{-\phi(z)}[-\frac{1}{4{g_5}^2}F_{mn}F^{mn}],
\end{equation}
where $F_{mn}=\partial_mA_n-\partial_nA_m$. $\phi(z)$ is a background scalar field that includes three energy parameters,
\begin{equation}\label{eq16}
  \phi(z)=m_b^2z^2+M_bz+\tanh(\frac{1}{M_bz}-\frac{m_b}{\sqrt{\Gamma_b}}).
\end{equation}
The three energy parameters $m_b$ expresses the bottom quark mass, $\Gamma_b$ means the string tension of the quark pair($b\bar{b}$) and $M_b$ represents a large mass related to the bottomonium non-hadronic decay. There exists a matrix element $\langle0|J_\mu(0)|\Upsilon(1S)\rangle=\epsilon_\mu f_n m_n$ ($\langle0|$ is the hadronic vacuum, $J_\mu$ is the hadronic current and $f_n$ indicates the decay constant) when the bottomonium decays into leptons. The value of three energy parameters for bottomonium in the background scalar field, provided by fitting the spectrum of masses in~\cite{Braga:2018zlu}, are respectively:
\begin{eqnarray}
  m_b=2.45GeV,\,\,\sqrt{\Gamma_b}=1.55GeV,\,\, M_b=6.2GeV.
\end{eqnarray}

\section{\label{sec:03} The differential configuration entropy}
With all the preparations in place, let's calculate the differential configuration entropy for heavy vector mesons. The quantity $\rho(\vec{x})$ in eq.~(\ref{eq4}) is the energy density which is the $T_{00}$ component of the energy-momentum tensor. The energy tensor from action \eqref{eq15} can be expressed in the following form with metric \eqref{eq10} :
\begin{equation}\label{eq19}
  T_{\mu\nu}:=-\frac{2}{\sqrt{-g}}\frac{\delta S_{\text{matter}}}{\delta g^{\mu\nu}}.
\end{equation}
Here, we consider gauge $A_z=0$, and the remaining components $A_\mu=A_\mu(p,z)e^{-i\omega t+ipx_3}$ which corresponds to plane waves propagating in the $x_3$ direction with momentum $p$. Futher, the field $A_\mu(p,z)$ can be written as $A_\mu(p,z)=\eta_\mu v(p,z)$ with the polarization vector $\eta_\mu$, where $v(p,z)$ is known as the bulk-to-boundary propagator corresponding to the quasinormal modes, satisfing $v(p,z)|_{z\rightarrow 0}=1$. The boundary value of the vector field $A_\mu(p,z)|_{z\rightarrow 0}=\eta_\mu$ can be seen as the source of the correlation functions of boundary current operator. In the zero temperature limit, the corresponding  solutions are called the normal modes representing the vacuum hadronic states. In gauge/gravity duality, quasinormal modes are normalizable gravity solutions denoting the quasi-particle states in a hot and dense medium. The real part of the corresponding complex frequency $\omega$, $Re(\omega)$, is interpreted as the effective mass and the imaginary part, $Im(\omega)$, is interpreted as the decay width. Finally, we write the vector field $A_\alpha$ in the gauge invariant electric field $E_\alpha=p_\alpha A_t+\omega A_\alpha$. Thus, in the zero momentum limit, the energy density reads
\begin{equation}\label{eq20}
  \rho(z)=T_{00}(z)=\frac{z^2e^{-\phi(z)}}{2g_5^2}(|E_\alpha|^2+\frac{af^2(z)}{|\omega|^2}|E_\alpha^\prime|^2).
\end{equation}
It is not difficult to find from the eq.~(\ref{eq20}) that quasinormal frequencies must be known to obtain the differential configuration entropy~\eqref{eq8}. Therefore, the quasinormal modes must be solved. 

First, we solve the equations of motion coming from eq.(\ref{eq15}) for the Gauss-Bonnet gravity  with the metric (\ref{eq10}). By taking the radial gauge $A_z=0$ and taking into account plane wave solutions $A_\mu(z,x_3,t)=e^{-i\omega t+iqx_3}A_\mu(z,\omega,q)$, the equations of motion are as follows
\begin{eqnarray}
  A_t''-(\frac{1}{z}+\phi')A_t'-\frac{q}{f}(qA_t+\omega A_{x_{3}}) &=& 0,\quad\quad\\
  A_\beta''+(\frac{f'}{f}-\frac{1}{z}-\phi')A_{\beta}'+\frac{1}{af^2}(\omega^2-aq^2f) )A_{\beta} &=& 0,\\
  A_{x_3}''+(\frac{f'}{f}-\frac{1}{z}-\phi')A_{x_3}'+\frac{\omega}{af^2}(qA_t+\omega A_{x_{3}}) &=& 0,\\
  \omega A_t'+aqfA_{x_3}' &=& 0 \label{eq24}.
\end{eqnarray}
where $\beta=(x_1,x_2)$ and the prime $(')$ denotes the derivative with respect to radial coordinate $z$. The vector field $A_\mu$ can be written in gauge invariant electric fields defined by $E_\mu$:  $E_\|=\omega A_{x_3}+qA_t$ and $E_{\bot}=\omega A_\beta$, where $E_\|$  indicates the propagation along the momentum direction and  $E_{\bot}$ means the propagation perpendicular to the momentum direction. So the corresponding electric field component equations read
\begin{equation}\label{eq25}
\begin{split}
0=&E_\bot''+(\frac{f'}{f}-\frac{1}{z}-\phi')E_\bot'+\frac{\omega^2-aq^2f}{af^2}E_\bot,\\
0=&E_\|''\!\!+\!\!(\frac{aq^2f^\prime}{\omega^2\!\!-\!\!aq^2f}\!+\!\frac{f'}{f}\!\!-\! \frac{1}{z}\!-\!\phi')E_\|'\!\!+\!\!\frac{\omega^2\!\!-\!\!aq^2f}{af^2} E_\|,
\end{split}
\end{equation}
One can easily find that the equations for both the longitudinal component and the transverse component are singular for $z=(0, z_h)$. To numerically solve the above electric field equations, we need two boundary conditions. To impose the boundary conditions at the event horizon, it is convenient to rewrite each of electric field equations into an Schr\"{o}dinger-like form by defining the tortoise coordinate, $\partial_{r_*}=-af(z)\partial_z$, and performing the Bogoliubov transformation, $E_\|=e^{\mathcal{B}_\|/2}\psi_\|\,, E_\bot=e^{\mathcal{B}_\bot/2}\psi_\bot$ with $\mathcal{B}_\|=\log \left(a q^2 f(z)-\omega ^2\right)+\phi (z)+\log (z)$ and $\mathcal{B}_\bot=\phi (z)+\log (z)$. The Schr\"{o}dinger-like equation is given by
\begin{equation}\label{eq26}
\begin{split}
\partial^2_{r_*}\psi_\|+(\omega^2-U_\|)\psi_\|&=0,\\
 \partial^2_{r_*}\psi_\bot+(\omega^2-U_\bot)\psi_\bot&=0,
 \end{split}
\end{equation}
with the effective potential
\begin{equation}\label{eq27}
\begin{split}
U_\|=&\frac{1}{4} (a^2 f(z)^2 \left(\frac{a q^2 f'(z)}{a q^2 f(z)-\omega^2}+\phi '(z)+\frac{1}{z}\right)^2
+a^2 f(z)\\
&\left(4 q^2-2 f'(z)\left(\frac{a q^2 f'(z)}{a q^2 f(z)-\omega ^2}+\phi'(z) +\frac{1}{z}\right)\right)+2 a^2\\
&f(z)^2 (\frac{a q^2\left(f''(z) \left(\omega ^2-a q^2 f(z)\right)+a q^2f'(z)^2\right)}{\left(\omega ^2-a q^2 f(z) \right)^2} +\frac{1}{z^2}\\
&-\phi''(z))-4 (a-1) \omega ^2 ),\\
U_\bot=&\frac{1}{4 z^2}(2 a^2 z f(z) \left(2 q^2 z-f'(z) \left(z \phi'(z)+1\right)\right)+a^2 f(z)^2\\
&\left(-2 z^2 \phi ''(z)+z^2 \phi'(z)^2+2 z \phi '(z)+3\right)-4 (a-1) \omega ^2 z^2),
\end{split}
\end{equation}
At the event horizon $U_{\|/\bot}(z_h)=0$, one can anlytically solve eq.\eqref{eq26} and get $\psi_{\|/\bot}(z_h)=c_1 e^{-i\omega r_*}+c_2 e^{+i\omega r_*}$, which corresponds to the superposition of incoming and outgoing waves. Because nothing can escape from the black hole,  only the incoming wave solution is physically allowed. Therefore, at $z\rightarrow z_h$, the incoming wave solution for Sch\"{o}dinger-like equation \eqref{eq26} has the following expansion:
\begin{equation}\label{eq28}
    \psi_{\aleph}=e^{-i\omega r_*}(1+a_{\aleph}(z-z_h)+b_{\aleph}(z-z_h)^2),
\end{equation}
with
\begin{equation}\label{eq29}
\begin{split}
    a_{\aleph}=&\frac{1}{M \omega ^4 z_h^4}(8 a \omega ^2 \text{sech}^2\left(\frac{1}{Mz_h}-\frac{\kappa}{\sqrt{\Gamma }}\right)-4 a M z_h \\
   &\left(8 a q^2\delta_{\aleph \|}+\omega ^2 \left(4 \kappa^2 z_h^2+2 M z_h+q^2 z_h^2+2\right)\right)),
   \end{split}
\end{equation}
where the formula of $b_\aleph$ is too long and will not be shown here and $\delta_{\aleph \|}$ is the Kronecker delta function with $\aleph=(\|,\bot)$. 

Then the incoming wave boundary condition can be obtained by the inverse Bogoliubov transformation as follows:
\begin{equation}\label{eq30}
\begin{split}
    E_{\aleph}(z_h)=&e^{-i\omega r_*(z_h)+\frac{\mathcal{B}_\aleph(z_h)}{2}},\\
    E^\prime_{\aleph}(z_h)=&(-i\omega r^{\prime} _{*} (z_h)+\frac{\mathcal{B}^\prime_\aleph(z_h)}{2}+a_\aleph)E_\aleph(z_h).
\end{split}
\end{equation}
Next, one can use eq.\eqref{eq30} as boundary conditions to numerically solve eq.\eqref{eq25} and search for complex
frequencies $\omega$ making $E(0)=0$. The real part of the corresponding quasinormal frequency is the effective mass, the imaginary part gives the decay width, and the corresponding solution is the quasinormal modes denoting the heavy quarkonium.

We can now turn to the differential configuration entropy. First of all, we perform a Fourier transform on the energy density $\rho(z)$ in the coordinate $z$ and split the energy density $\rho(k)$ as follows
\begin{eqnarray}
%\nonumber to remove numbering (before each equation)
 \rho(k) &=& (C(k)-iS(k))/\sqrt{2\pi},\nonumber \\
  C(k) &=& {\int_0}^{z_h}\rho(z)\cos(kz)dz,\label{eq31} \\
  S(k) &=& {\int_0}^{z_h}\rho(z)\sin(kz)dz,\nonumber
\end{eqnarray}
with the modal fraction,
\begin{equation}\label{eq32}
  \epsilon(k)=\frac{S^2(k)+C^2(k)}{[S^2(k)+C^2(k)]_{max}}.
\end{equation}
So the differential configuration entropy reads
\begin{equation}\label{eq33}
  S_{DCE}=-\int_{-\infty}^{\infty}\epsilon(k)\log[\epsilon(k)]dk.
\end{equation}

\section{\label{sec:04} The Results of effective potential }
\begin{figure}
  \centering
  \includegraphics[width=7cm]{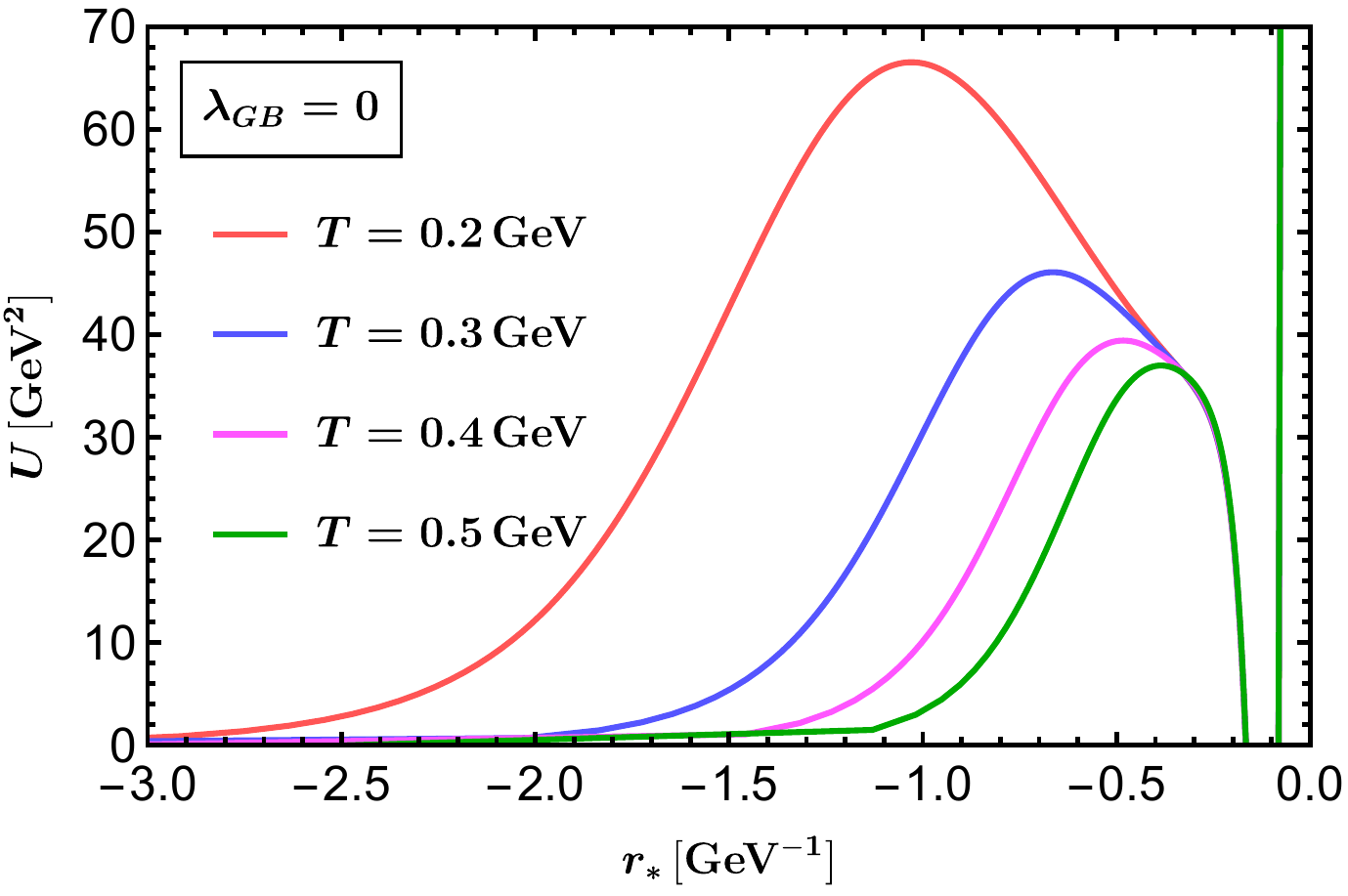}
  \caption{ The effective potential as a function of the tortoise coordinates with varying temperatures. From top to bottom, the curve sequentially represents $T=0.2,0.3,0.4,0.5 \,\text{GeV}$. }\label{fig2}
\end{figure}
\begin{figure}
  \centering
  \includegraphics[width=7cm]{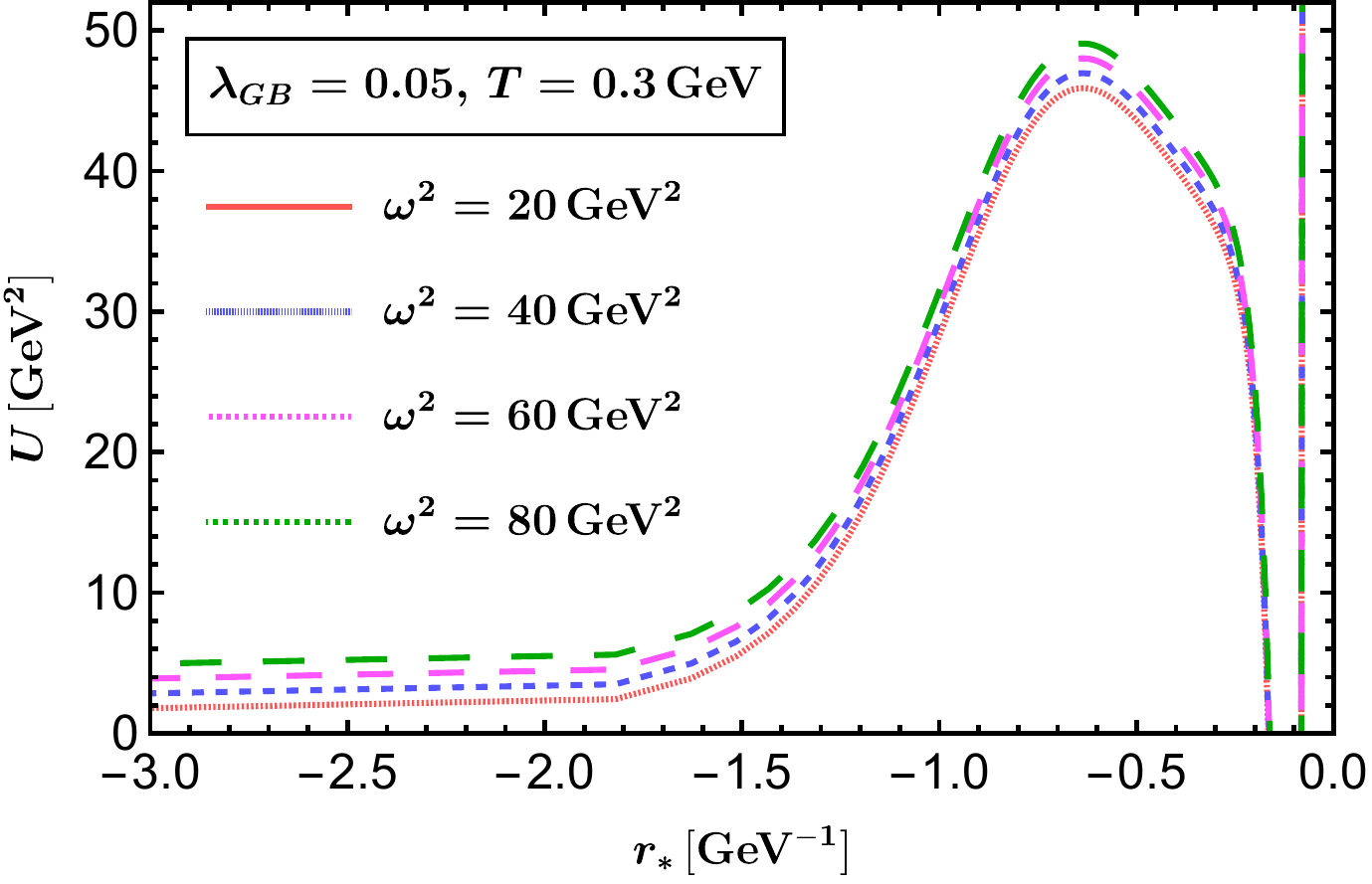}
  \includegraphics[width=7cm]{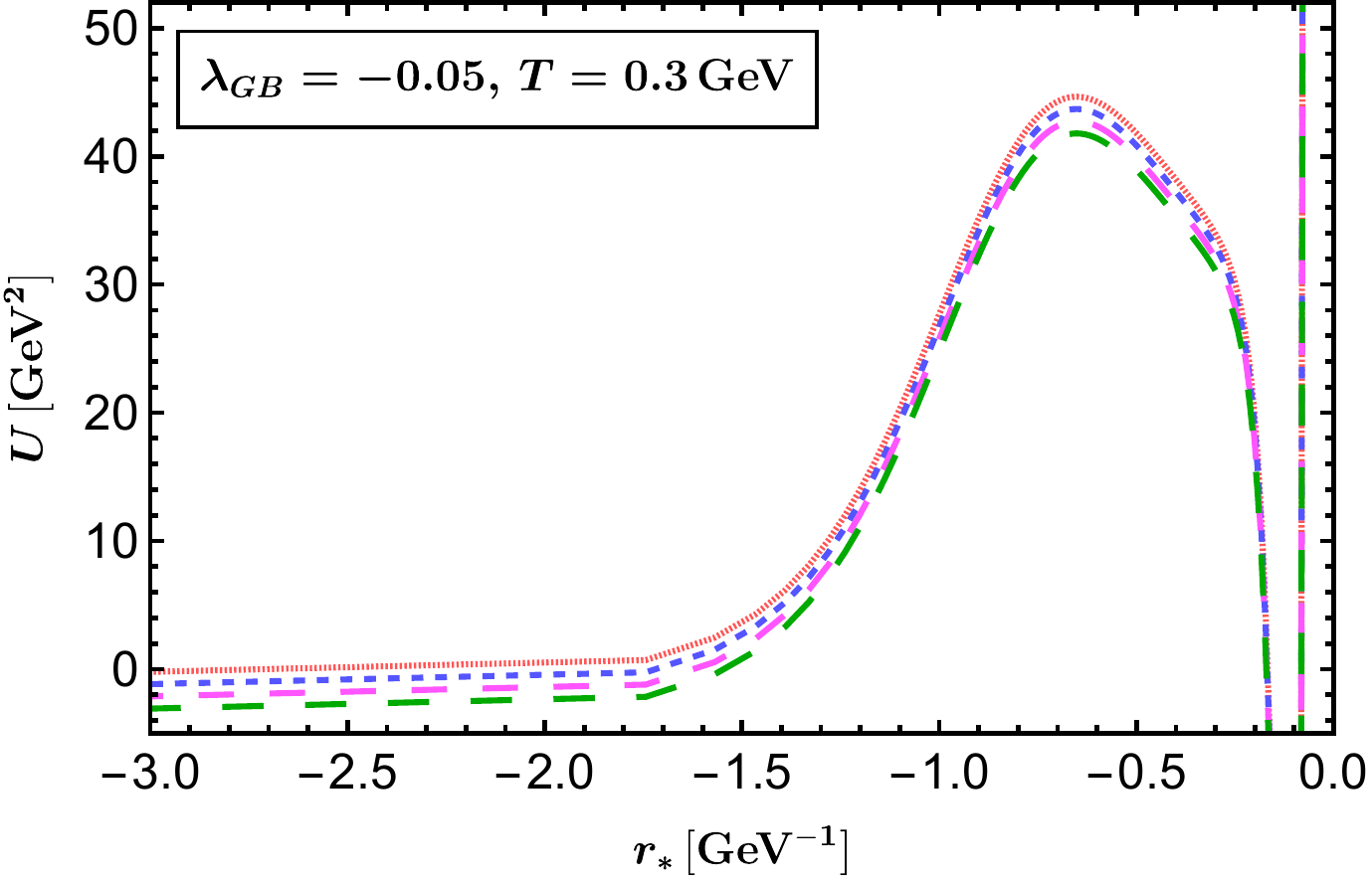}
  \caption{ The effective potential as a function of the tortoise coordinates with varying energies at $\lambda_{GB}=0.05$ (top panel) and $\lambda_{GB}=-0.05$ (bottom panel).  The red, blue, magenta, green curve sequentially denotes  $\omega ^2=20,40,60,80 \,\text{GeV}^2$. }\label{fig3}
\end{figure}
Let us start by studying the effective potential for vanishing wave number $q=0$, for which the effective potential $U_\|$ and $U_\bot$ in Eq.\eqref{eq27} have the same form,
\begin{equation}\label{eq34}
\begin{split}
    U=&U_\||_{q=0}=U_\bot|_{q=0}=\frac{1}{4 z^2}\\
    &(-2 a^2 z f(z) f'(z) \left(z \phi '(z) +1\right)+a^2 f(z)^2\\
    & \left(-2z^2 \phi ''(z)+z^2 \phi '(z)^2+2 z \phi '(z)+3\right)\\
    &-4 (a-1) \omega ^2 z^2).
\end{split}
\end{equation}
One can find that the behavior of effective potential caused by higher-order curvature corrections is determined by the last term $(1-a) \omega ^2$, which results in $U(z_h)\neq 0$ at the horizon for $a\neq 1$ or $\lambda_{GB}\neq 0$ and the effective potential depends on the energy $\omega$. The effective potential as a function of the tortoise coordinates with varying temperature is shown in  Fig.\ref{fig2}. The potential shape behaves as a well which suggests some quasiparticle states in the dual theory exist in the range of temperature, corresponding to the bottomonium in the hot QGP. The depth of potential well decreases with the increase in temperature, indicating the dissociation effect of bottomonium increases when the temperature increases. In addition, near the boundary($r_*=0$), there exists an infinite barrier localized which starts from negative potential.

The appearance of higher-order curvature corrections also causes the effective potential depending on the squared energy $\omega^2$. The effect of squared energy on effective potential is related to the temperature parameter $a$. When $a \leq 1$ ($\lambda_{GB} \geq 0$), the depth of potential well increases as the squared energy increases, which means the probability of bottomonium formation increases with the increasing energy, as displayed in the upper panel of Fig.\ref{fig3}. When $a \geq 1$ ($\lambda_{GB} \leq 0$), the depth of potential well decreases as the squared energy increases, which indicates the dissociation effect of bottomonium increases with the increasing energy, as displayed in the lower panel of Fig.\ref{fig3}. 

\begin{figure}
  \centering
  \includegraphics[width=7cm]{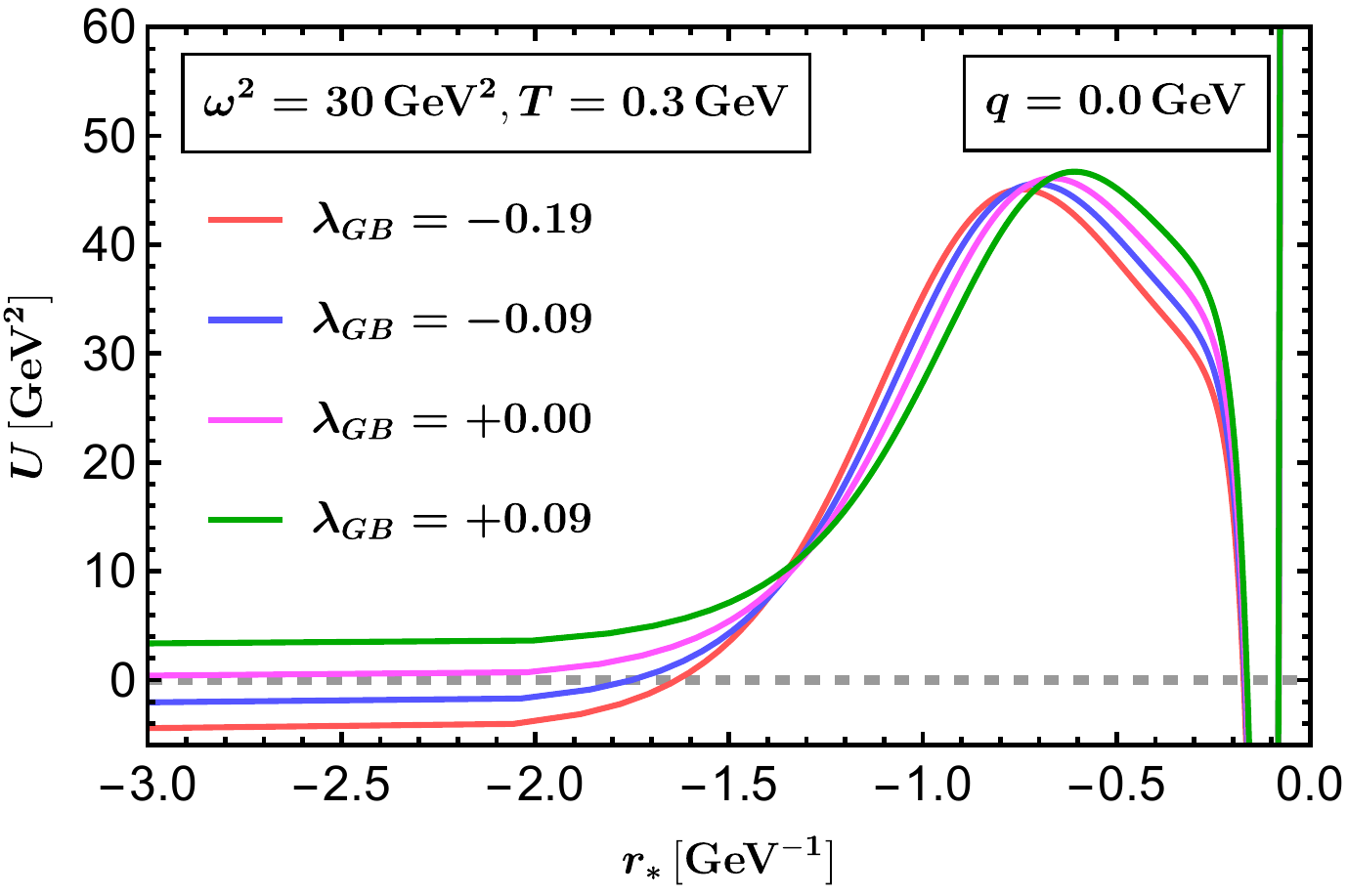}
   \includegraphics[width=7cm]{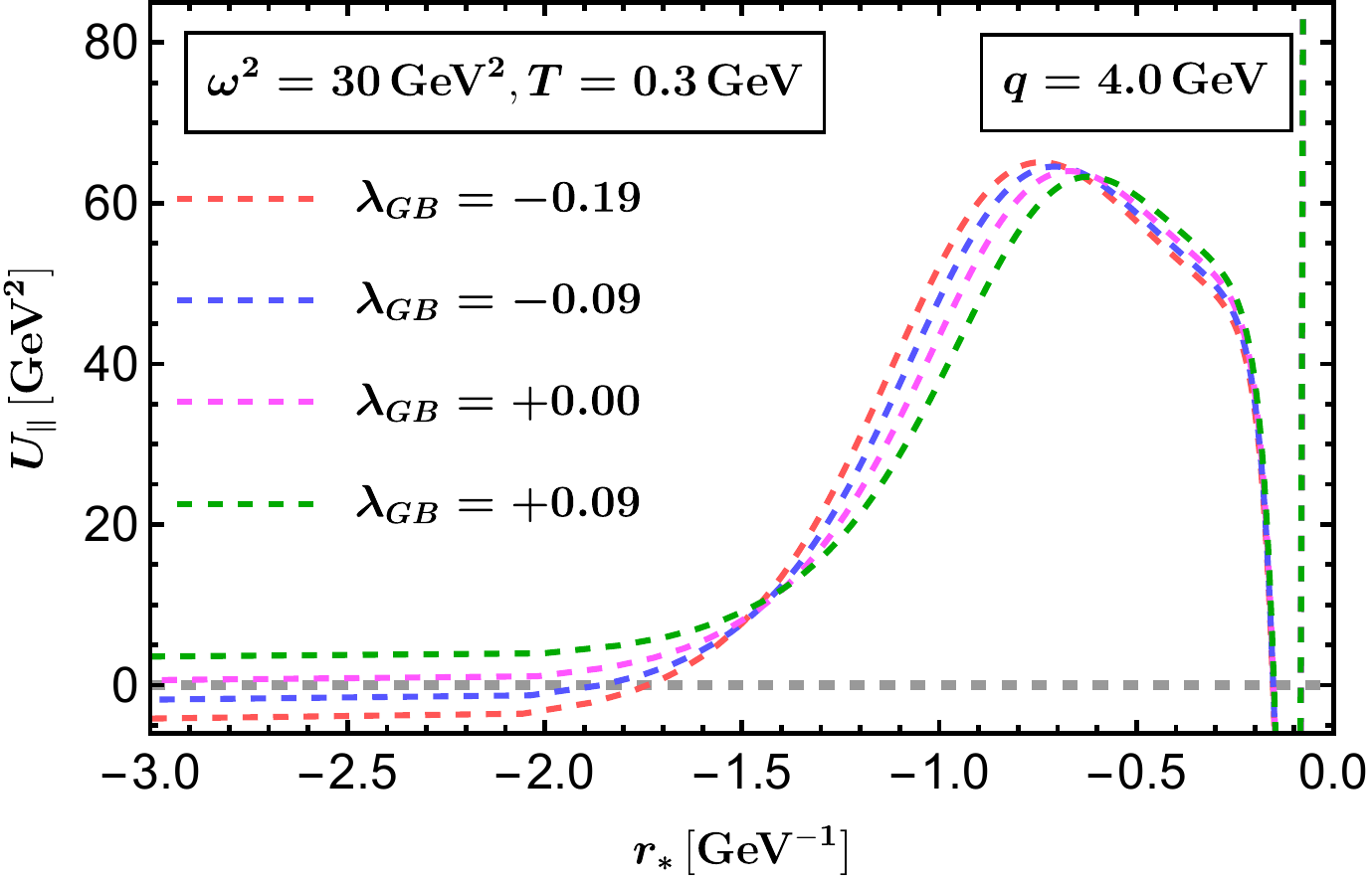}
    \includegraphics[width=7cm]{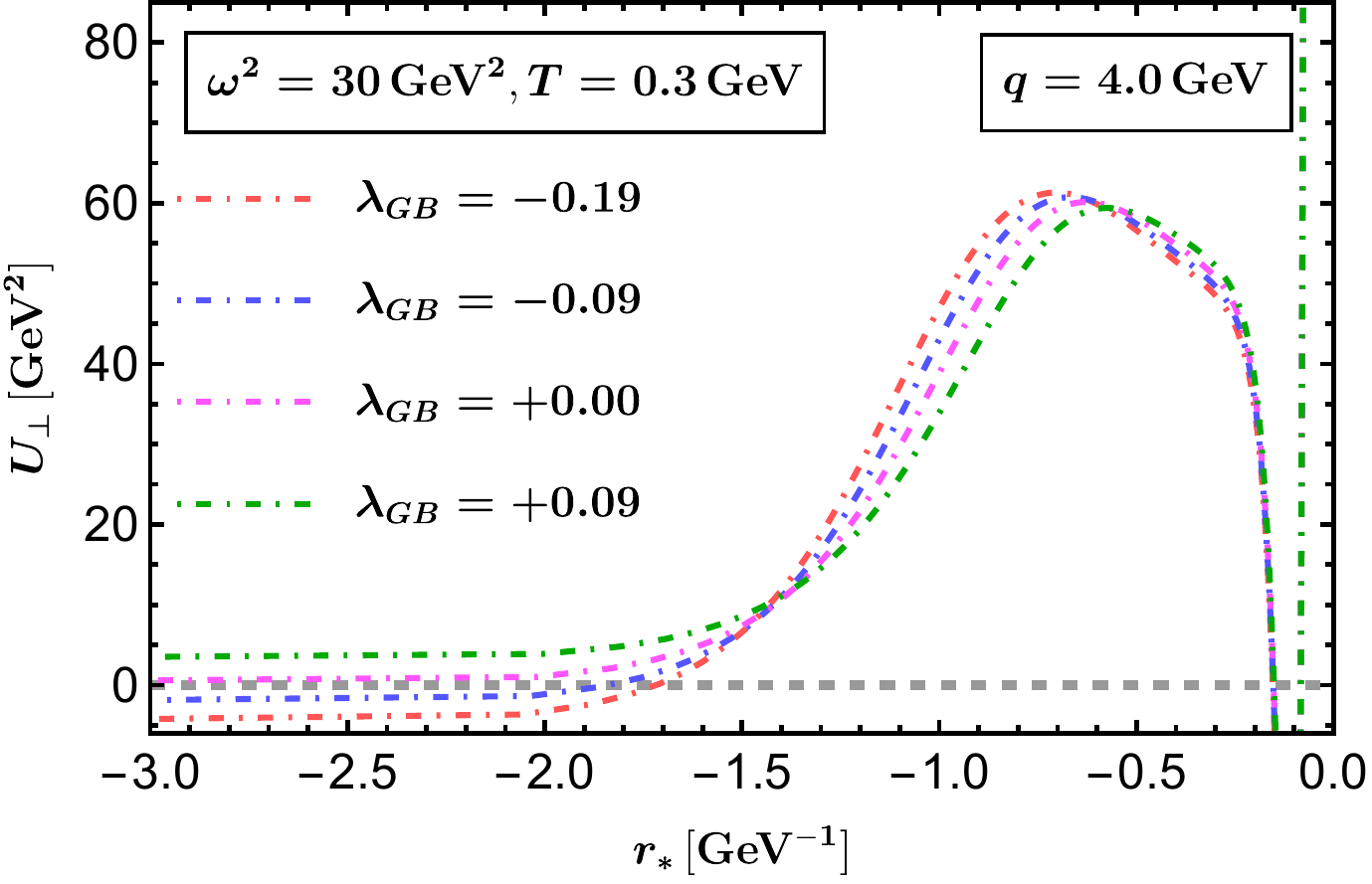}
  \caption{ The effective potential as a function of the tortoise coordinates with different $\lambda_{GB}$ at zero wave number and finite wave number. The red, blue, magenta, and green curve sequentially denotes  $\lambda_{GB}=-0.19,-0.09,0.0,0.09$ on the upper($q=0.0 \,\text{GeV}$), middle($q=4.0 \,\text{GeV}$ for the parallel case), and lower($q=4.0 \,\text{GeV}$ for the perpendicular case) panels. }\label{fig4}
\end{figure}

In Fig.\ref{fig4}, we plot the dependence of effective potential as a function of tortoise coordinate on the constant $\lambda_{GB}$. An interesting phenomenon is, as increasing constant $\lambda_{GB}$, the potential well becomes deeper for zero wave number, corresponding to the probability of bottomonium formation being larger, while that becomes shallower for finite wave number indicating the melting effect being stronger. This behavior looks very unreasonable. Because the large $\lambda_{GB}$ means the coupling strength between plasma and heavy quarks is larger, which indicates the dissociation effect is stronger for large $\lambda_{GB}$. We explain this non-trivial behavior by the interplay of the interaction between the two heavy quarks and the interaction between the medium with each of the heavy quarks\cite{Zhao:2021ogc}. One can well understand that a $Q\bar{Q}$ bound state denotes a balance between repulsive kinetic( by an entropic effect) and attractive potential energy(by an internal energy). Color screening modifies the attractive binding energy between the quarks, while the increasing entropy with large separation gives rise to a growing repulsion\cite{Satz:2015jsa}.  For smaller wave numbers, the interaction(binding) is stronger between quark and anti-quark. However, for larger wave numbers, the interaction(dissociation) is stronger between quark/anti-quark and mediums.  The non-trivial behavior suggests that here a two-body potential approach is no longer applicable, so a kind of method related to an enhanced collective entropic quarkonium dissociation may be used\cite{Kharzeev:2014pha}, such as configuration entropy. Notes that, at $\lambda_{GB}<0$, the effective potential at the horizon $U(z_h)$ is less than $0\text{GeV}^2$, however, which is greater than $0\text{GeV}^2$ for $\lambda_{GB}>0$ .

\section{\label{sec:05} The Results of QNMs and differential configuration entropy }
\begin{figure}
  \centering
  \includegraphics[width=7cm]{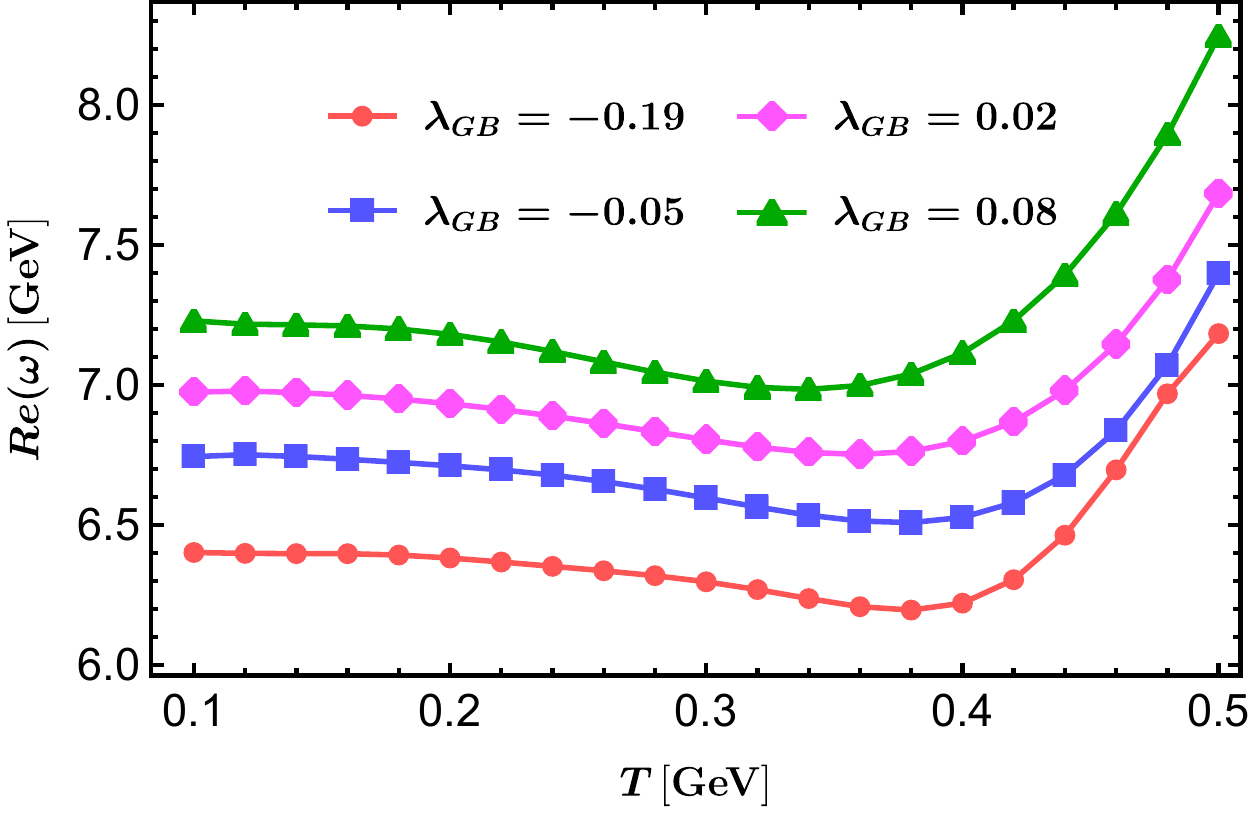}
  \caption{ The real part of the quasinormal frequencies as a function of the temperature $T$ with different constants $\lambda_{GB}$. From bottom to top, the curve represents $\lambda_{GB}=-0.19,-0.05,0.02,0.08$, respectively. }\label{fig5}
\end{figure}
\begin{figure}
  \centering
  \includegraphics[width=7cm]{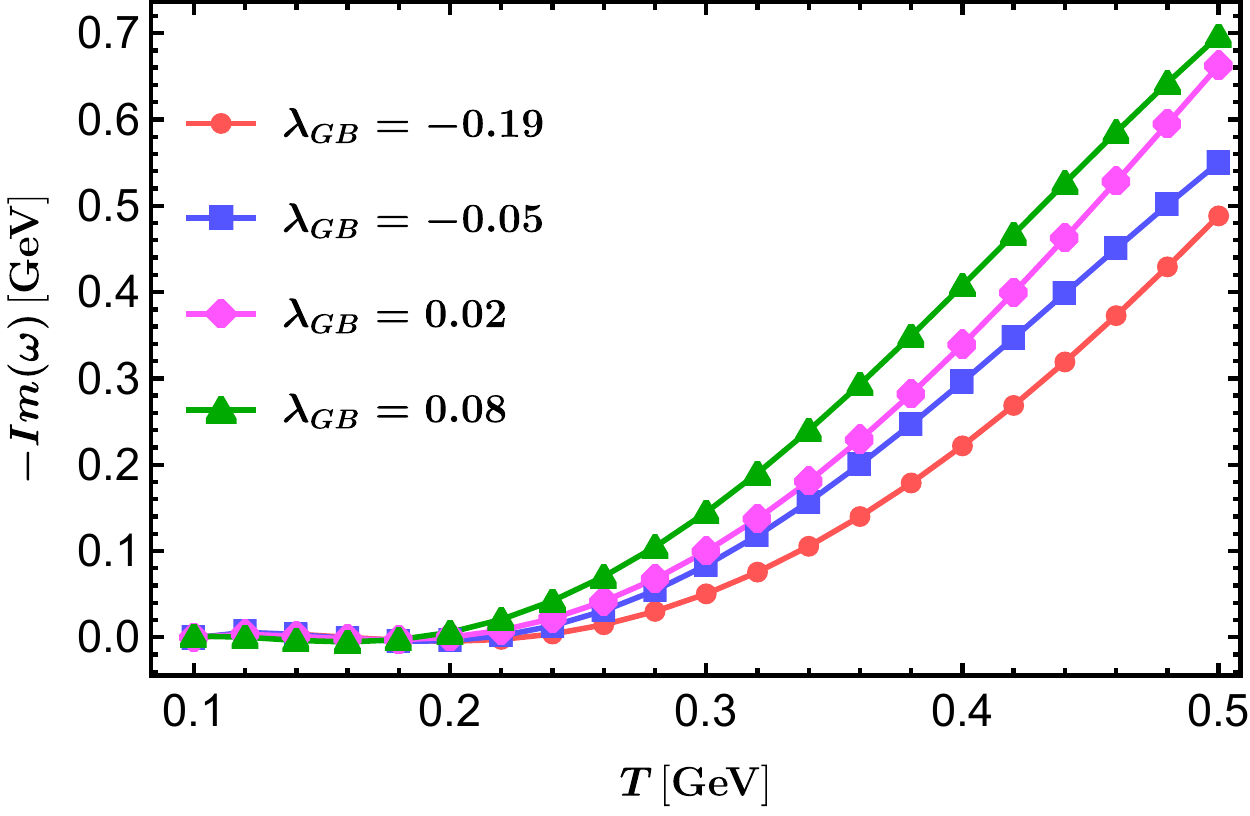}
  \caption{ The imaginary part of the quasinormal frequencies as a function of the temperature $T$ with different constants $\lambda_{GB}$. From bottom to top, the curve represents $\lambda_{GB}=-0.19,-0.05,0.02,0.08$, respectively. }\label{fig6}
\end{figure}
\begin{figure}
  \centering
  \includegraphics[width=7cm]{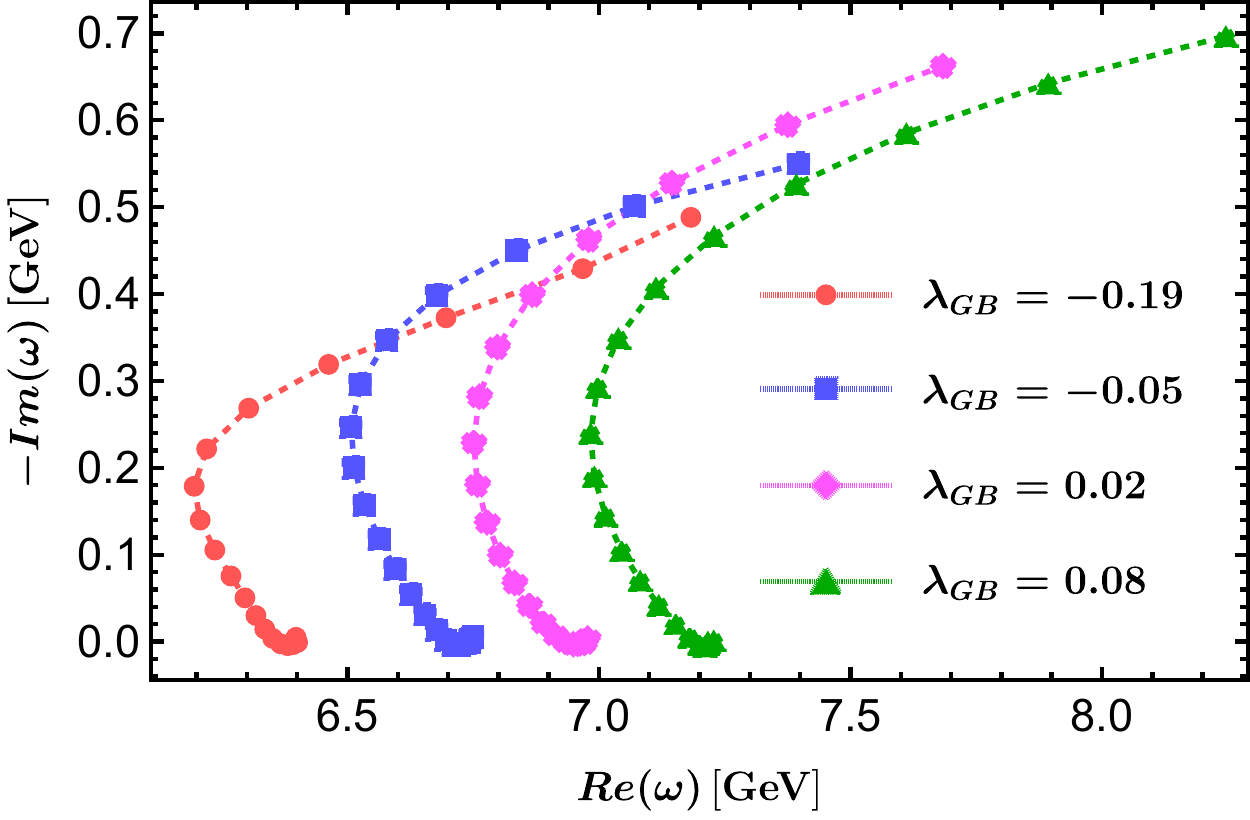}
  \caption{ Quasinormal spectrum of electric field fluctuations in the plane of complex $\omega$ defined by Eq.\eqref{eq25}. From left to right, the curve represents $\lambda_{GB}=-0.19,-0.05,0.02,0.08$, respectively.}\label{fig7}
\end{figure}
\begin{figure}
  \centering
  \includegraphics[width=7cm]{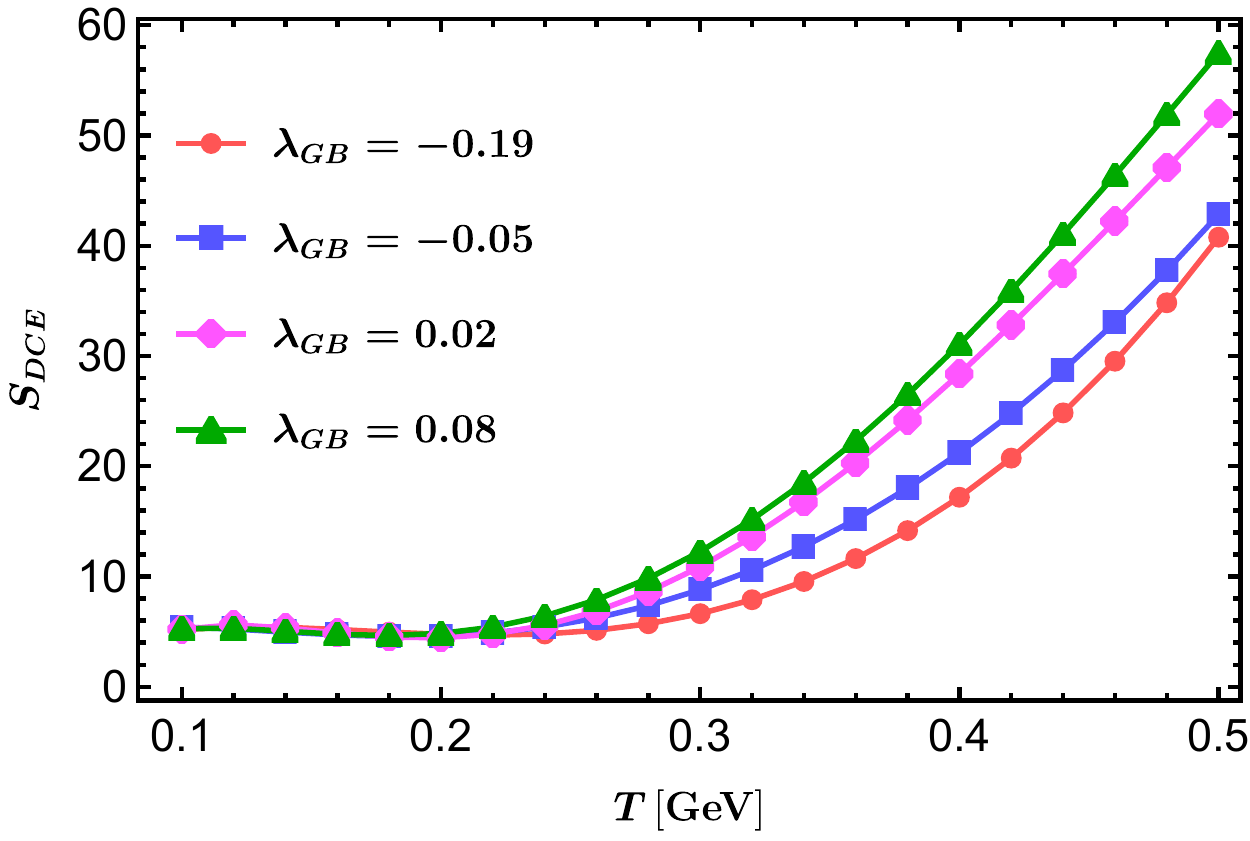}
  \caption{ The dependence of the differential configuration entropy on temperature with varying constants $\lambda_{GB}$. From bottom to top, the curve represents $\lambda_{GB}=-0.19,-0.05,0.02,0.08$, respectively. }\label{fig8}
\end{figure}

Quasinormal spectra (QNMs) is widely used in different research fields to study different physics. In holography, research of QNMs became a standard tool in treating the near-equilibrium problem of gauge theory. It gives the location of the poles of the retarded correlators in the gauge theory, which provides worthwhile information about the quasiparticle spectra function and some transport coefficients. 

Here, we consider the heavy mesons at rest in the hot medium. The real part of the quasinormal frequencies is interpreted as the effective mass and the imaginary part describes the decay width. In Fig.\ref{fig5}, we plot the dependence of effective mass on the temperature at four different values of the constant $\lambda_{GB}$ and the decay width is presented in Fig.\ref{fig6}. One can find that the effective mass monotonically decreases with temperature in the lower temperature region, while an opposite behavior is obtained in the higher temperature region, which is consistent with Ref. \cite{Zhao:2021ogc, Braga:2019xwl}. In addition, the effective mass increases with the increase of constant $\lambda_{GB}$. As expected, Fig.\ref{fig6} shows the decay width (the reciprocal of lifetime) decreases with reduced temperature and the dissociation effect becomes stronger with the increase of $\lambda_{GB}$. The QNMs reveal the relation between effective mass and lifetime of quasiparticle state in different temperature ranges, as displayed in Fig.\ref{fig7}. For each curve, the temperature increases from bottom to top. One can find that the dependence of quasi-particle lifetime on effective mass is stronger and the strength of the dependence is basically independent of constant $\lambda_{GB}$ in the lower temperature region. However, in the higher temperature region, the sensitivity of quasi-particle lifetime on effective mass increases with the increase of constant $\lambda_{GB}$.

Finally, we calculate the DCE in function of temperature with different constant $\lambda_{GB}$ in Fig.\ref{fig8}. The DCE becomes bigger with the increasing temperature, suggesting the dissociation effect being stronger, which is the same as Ref.\cite{Braga:2020hhs}. Larger $\lambda_{GB}$ also increases the DCE in the higher temperature region, this result is reasonable since larger $\lambda_{GB}$ (or smaller $\eta/s$) in general means stronger coupling, which implies that the interaction between quark/anti-quark and medium is stronger than that between quark and anti-quark. In other words, larger $\lambda_{GB}$ should have a stronger dissociation effect corresponding to a larger DCE. However, the DCE basically remains unchanged with varying temperature and $\lambda_{GB}$ in the lower temperature regions, which means the existence of bottomonium is not affected by the thermal medium.

\section{\label{sec:06} Conclusion and discussion }

In this letter, we study the effective potential, quasinormal spectra (QNMs) and  the differential configuration entropy (DCE). The results show that the effective potential is not a good physical quantity to estimate the dissociation strength of bottomonium under the very strong coupling QGP. At the moment,  a kind of method related to an enhanced collective entropic quarkonium dissociation may be used, such as configuration entropy. By calculating the QNMs, we find the behavior of effective mass as a function of temperature is non-monotonic and the effective mass is larger for strong coupling plasma. In particular, the results from QNMs and DCE indicate both higher temperature and stronger coupling promote the dissociation effect, leading to the suppression of the final yield from dilepton pair.

With some interesting conclusions obtained from this paper, we are also curious about the properties of $J/\Psi$ under the strongly coupled plasma. We will do this job in the future. In addition, we guess the effect of strong coupling may affect the heavy vector mesons' spectra, which is directly related to the production rate of dilepton. One can calculate the current-current correlation function of heavy quark to obtain it and compare the result with the experiment, which may give a surprising conclusion. Of course, there are also many other physical phenomena in experiments that urgently require a self-consistent theory to explain, such as spin alignment. One can also study the effect of rotating QGP on the QNMs and the DCE, as demonstrated in Ref.\cite{Zhao:2022uxc}.

%\paragraph{Note added.}
\section{acknowledgments}

This work is supported in part by the National Key Research and Development Program of China under Contract No. 2022YFA1604900. This work is also partly supported by the National Natural Science Foundation of China (NSFC) under Grants No. 12275104, No. 11890711, No. 11890710, and No. 11735007, as well as by the Fundamental Research Funds for the Central Universities (Innovation Funding Projects) 2020CXZZ108.
%\appendix

\clearpage

\section*{Reference}
\bibliography{ref}

\end{document}